\documentclass[12pt]{article}

\usepackage{graphicx}
\usepackage{amsmath}
\usepackage{amssymb}
\usepackage{fancyhdr}
\usepackage{color}
\newcommand{\be}{\begin{equation}}
\newcommand{\ee}{\end{equation}}
\newcommand{\bea}{\begin{eqnarray}}
\newcommand{\eea}{\end{eqnarray}}

\newcommand{\expect}[1]{\left\langle#1\right\rangle}

\newcommand{\intdc}{\int\!{\rm d}\vec{c}\,}
\newcommand{\intdrp}{\int\!{\rm d}\vec{r}_n\int\!{\rm d}\vec{p}_n\,}
\newcommand{\intdrph}{\int\!{\rm d}\vec{r}\int\!{\rm d}\vec{p}/h^3\,}

\newcommand{\Prt}{\widetilde{\Pr}}

\date{}
\begin{document}
\title{{\bf The Brandeis Dice Problem \& Statistical Mechanics}}
\author{Steven J. van Enk\footnote{email:
{\tt svanenk@uoregon.edu}}\\
Department of Physics, University of Oregon,\\
 Eugene, OR 97403, USA
}

\maketitle
Jaynes invented the Brandeis Dice Problem as a simple illustration of
the MaxEnt (Maximum Entropy) procedure that he had demonstrated to work so well in Statistical Mechanics.
I construct here two alternative solutions to his toy problem. One, like Jaynes' solution, uses MaxEnt and yields an analogue of the canonical ensemble, but at a different level of description. The other uses Bayesian updating and
yields an analogue of the micro-canonical ensemble.
Both, unlike Jaynes' solution, yield error bars, whose operational merits I discuss.
These two alternative solutions are not equivalent for the original Brandeis Dice Problem, but become so in what must, therefore, count as the analogue of the thermodynamic limit, $M$-sided dice with $M\rightarrow\infty$.
Whereas the mathematical analogies between the dice problem and Stat Mech are quite close, there are physical properties 
that the former lacks but that are crucial to the workings of the latter. Stat Mech is more than just MaxEnt.
\\


\newpage
\section{Introduction}
Jaynes introduced the following problem in his 1962 lectures for the Brandeis Summer School (the underlining is his):
\begin{verse}
A die has been tossed a very large number $N$ of times, and we are told that the \underline{average} number of spots per toss was not 3.5, as we might expect from an honest die, but 4.5. Translate this information into a probability assignment $P_n$, $n=1,2,\ldots,6$, for the $n$-th face to come up on the next toss.
\end{verse}
Jaynes then went on to solve this problem---now known as the Brandeis Dice Problem---by using his celebrated\footnote{For example, the MaxEnt workshop is, as of 2014, in its 34th incarnation.} MaxEnt procedure. That is, he 
maximized the entropy $S=-\sum_n P_n\log P_n$, subject to the trivial constraint $\sum_nP_n=1$ and the less trivial
$\sum_n nP_n=\expect{B}$, where $\expect{B}$ denotes the observed average number of spots (here, $\expect{B}=4.5$). The numerical solution thus obtained looks quite reasonable at first sight (see the first two columns of Table 1), and it seems clear that Jaynes intended  this problem as an innocent little example of the procedure that he had applied to 
the more serious problems of Statistical Mechanics earlier
in 1957. In particular, where it was known that the canonical ensemble arises from maximizing the Gibbs entropy, subject to a constraint on the expected value of the total energy, Jaynes' novel epistemic interpretation and justification of this procedure were that it takes into account what we know (the constraint) but nothing more. Lack of knowledge is quantified properly by Shannon's entropy, which in turn equals (not just numerically but, on Jaynes' viewpoint, also conceptually) the Gibbs entropy.  The close relation between Jaynes' solution to the Brandeis Dice Problem and the canonical ensemble
is exemplified by the fact that his solution may be given in the form 
\be
P_n=\frac{ \exp(-\lambda n)}{Z(\lambda)},
\ee
with the ``partition function'' given by $Z(\lambda)=\sum_n \exp(-\lambda n)$,
and with the term $\exp(-\lambda n)$ being the analogue of the Boltzmann factor.
The Lagrange multiplier $\lambda$ is fixed by the constraint $d\ln Z(\lambda)/d\lambda=-\expect{B}$, and $\lambda$ can be positive or negative or zero, depending on whether
$\expect{B}$ is smaller than, larger than, or equal to 3.5 (the ``honest value'').

It turned out, however, that the Brandeis Dice Problem was not so innocent after all, and many criticisms were leveled against Jaynes' solution (see, for example, Rowlinson 1970 and Friedman and Shimony 1971). Although I will address one particular criticism below, it is not the main purpose of this article to discuss these criticisms in any detail or to defend Jaynes' solution against them---Jaynes is an  entertaining defender of his own principles, see Jaynes 1978, 1985.\footnote{Not to mention that I agree with {\em some} criticisms, especially those of Skyrms (1985) and Uffink (1996).} Rather, it is to point out that an alternative analogy of the canonical ensemble can be obtained by applying the MaxEnt procedure at a different level of description, involving a different sort of probabilities. 
That is, where Jaynes introduced degrees of belief about the die coming up $n$ spots for $n=1\ldots 6$,
one can introduce an underlying objective probability [aka chance] for the die to come up $n$ spots, and subsequently introduce degrees of belief about those chances. One can then apply the MaxEnt procedure to the latter and thus obtain a second analogue of the canonical ensemble.
One argument in favor of this more involved procedure is that 
Bayesian updating would, in fact, be applied at that same level. Moreover, it turns out that Bayesian updating then leads (in the limit $N\rightarrow\infty$) to the analogue of the micro-canonical ensemble. As far as I know this relation via ensembles between the MaxEnt and Bayesian updating procedures has not been noted before.

The two different types of probabilities that I propose to exploit here, one objective chance, the other a subjective degree of belief, are sometimes referred to as first-order and second-order probabilities, respectively (see, e.g., Kyburg 1988). This distinction was already made by Laplace, a fact that has not always been appreciated (for a nice discussion of Laplace's original writings, and much more about probabilities of various kinds, I refer the reader to (Maher, 2010)). I mention Laplace just because there is one criticism that has been raised against Jaynes' method as well as against Laplace's Principle of Indifference and his Rule of Succession (see Zabell 1989; see also the Appendix):  that it ought to be impossible to obtain  any  estimates of probabilities---let alone {\em precise} ones---from a mere lack of knowledge. As a byproduct, this paper will show how
Laplace and Jaynes could have preempted, at least partially, this particular criticism, if they had provided error bars (i.e., standard deviations) with their estimates. A few hundred years late, I will provide those missing error bars and show that they are quite large, of the same order of magnitude as the estimated probabilities themselves. Thus, these estimates are not precise at all.
This, then, is
one benefit of introducing both first-order and second-order probabilities: the latter come automatically equipped with standard deviations. 
I will discuss in some detail why these standard deviations are operationally useful in Section \ref{SD} below.

\section{Two alternative solutions}
I consider in this Section two alternative ways of interpreting the Brandeis Dice Problem, and show how each interpretation leads to a solution that differs from Jaynes' original solution. One of these two solutions is based on Bayesian updating. While it has been discussed before by Uffink in 1996 (see also Porto Mana 2009 for an extensive numerical investigation of various alternative Bayesian updating procedures for this problem, as well as the role played by different priors), my presentation emphasizes different aspects.
The other solution is new, as far as I know, and I will start with deriving it.
\subsection{...interpreted as imposing the constraint...}\label{ME}
The Brandeis Dice Problem concerns
what a Bayesian agent can say about her degrees of belief in the (truth of) the following 6 propositions for  $n=1,\ldots 6$:
\begin{verse}
$H_n$: On the next throw,  this particular die will show $n$ spots.
\end{verse}
Let us denote these 6 degrees of belief by Pr($H_n$ is true)$=:P_n$, and assume no other outcomes are considered by our agent, so that $\sum_n P_n=1$.

When Jaynes presented his MaxEnt solution he formulated the problem as follows (where I change his notation to mine)
\begin{verse}
Let us see what solution the Principle of Maximum Entropy gives for this problem, if we interpret the data as imposing the mean value constraint
\[
\sum_n nP_n=4.5.
\]
\end{verse}
Let us for the moment accept this particular way of interpreting the problem, namely, as imposing a constraint on the probabilities $\{P_n\}$. (I return to this assumption in the next subsection.) 

The probabilities $\{P_n\}$ are subjective degrees of belief about propositions, according to Jaynes. But we may in addition introduce objective probabilities, or chances, in two different ways, first assuming that those chances are real (i.e., physical), or, second, assuming they are merely convenient mathematical entities.
First, then, we could imagine that the die in question is going to be thrown by a precisely tuned mechanical device. Given  how the mechanical device tosses the sturdy die (with initial velocities of the die distributed, say, according a very narrow normal distribution) it seems we could introduce a possibly calculable chance that the outcome of any toss will be $n$ spots. Denote this chance by $c_n$. Our agent may then assign a probability (degree of belief) distribution over $\vec{c}:=(c_1,c_2,\ldots c_6)$, which we denote by $\Prt(\vec{c})$. It is normalized by  $\intdc \Prt(\vec{c})=1$,
and the relation between $P_n$ and $\Prt(\vec{c})$ is
\be
P_n=\intdc c_n \Prt(\vec{c}).
\ee
Now a radical anti-objectivist, like De Finetti (see his book from 1974), will deny the existence of $\vec{c}$ as a real thing. But even so, if he agrees that the sequence of tosses is exchangeable and extendible, then the probability of
finding exactly $M_n$ times the result $n$ (for $n=1\ldots 6$) in $N$ tosses is, according to the De Finetti theorem, still of the form
\be\label{DF}
\Pr(M_1,M_2,\ldots M_6)=\intdc\Prt(\vec{c})
\frac{N!}{M_1!M_2!\ldots M_6!} 
c_1^{M_1}\times
c_2^{M_2}\ldots \times c_6^{M_6},
\ee
{\em as if} the underlying chances exist. Both ways of introducing the chances $\vec{c}$ suffice for the purposes of this Section. For making the {\em physical} analogy with Statistical Mechanics, as opposed to the purely {\em mathematical} analogies pursued here,  stronger,  the first, objective, description may be preferable. See Section \ref{Stat Mech}.

Jaynes' constraint is now a constraint on the probability distribution $\Prt$:
\be\label{Prtc}
\intdc \sum_n nc_n\Prt(\vec{c})=\sum_nnP_n=\expect{B}.
\ee
Accepting this, it is now obvious that the MaxEnt principle ought to be applied to $\Prt$.
The well-known obstacle in the way of straightforwardly translating this principle into a condition on $\Prt$ is the necessity of having to choose a measure $m(\vec{c})$ on the space of the continuous variables $\vec{c}$.\footnote{Jaynes' solution avoids this problem by focusing on just discrete degrees of freedom, and it may, therefore, be considered as the most straightforward solution of the toy problem. However, as Seidenfeld (1986) pointed out, even discrete variables are not immune from the possibility of alternative descriptions: in this case one might consider the additional 1 or 2 sides of the die that are visible after the toss, and apply MaxEnt to the larger number of degrees of freedom that correspond to the more detailed description.
Even though the additional degrees of freedom may seem irrelevant, the MaxEnt solution thus obtained is different.} 
That is, we ought to optimize the {\em relative} entropy
\be
\widetilde{S}:=-\intdc \Prt(\vec{c})\ln\left[\frac{\Prt(\vec{c})}{m(\vec{c})}\right],
\ee
subject to normalization and the constraint (\ref{Prtc}).
I will choose here the natural looking $m(\vec{c})=$constant, and return to this choice when discussing analogies and disanalogies with Statistical Mechanics in Section \ref{Stat Mech}. 

The solution to the problem at hand is of the form
\be\label{Can1}
\Prt(\vec{c})=\frac{\exp(-\lambda\sum_n nc_n)}{\widetilde{Z}(\lambda)},
\ee
with the ``partition function'' given by
\be\label{Can2}
\widetilde{Z}(\lambda):=\intdc\exp(-\lambda\sum_n nc_n),
\ee
and $\lambda$ fixed by the transcendental equation 
\be
\frac{d\ln \widetilde{Z}(\lambda)}{d\lambda}=-\expect{B}.
\ee
(Just as for Jaynes' solution, the sign of $\lambda$ is easily seen to be the same as that of $3.5-\expect{B}$.)
The numerical solution for $\expect{B}=4.5$ is tabulated in Table 1 and may be compared to Jaynes' solution. Even though both solutions use the same constraint and both maximize the missing information, they differ, because they employ different levels of description.
\begin{table}[h]
\begin{center}
\begin{tabular}{|c|c|c|c|}
\hline
\hline
Probability& Jaynes&Chances &(Error bar)\\
\hline
\hline
$P_1$&0.054&0.068& (0.069)\\
\hline
$P_2$&0.079&0.083& (0.083)\\
\hline
$P_3$&0.114&0.106 &(0.103)\\
\hline
$P_4$&0.165&0.145& (0.133) \\
\hline
$P_5$&0.240&0.218& (0.176)\\
\hline
$P_6$&0.347&0.380 &(0.214)\\
\hline
\hline
\end{tabular}
\end{center}
\label{default}
\caption{Jaynes' MaxEnt solution to the Brandeis Dice Problem (second column) compared with an alternative solution (third column) calculated in Section \ref{ME} that applies the MaxEnt procedure to degrees of belief about the chances to throw $n$ dots.
The entry headed ``(Error bar)'' gives the standard deviation $\sigma_n$ in this alternative estimate of $P_n$, as calculated in Section \ref{SD}.}
\end{table}
\subsection{Bayesian Updating}\label{BU}
As both Skyrms (1985) and Uffink (1996) have argued, there seems to be no particular reason why one would interpret Jaynes' formulation of the Brandeis Dice Problem as giving rise to a constraint on the probability distribution $\{P_n\}$. It seems, within Bayesianism, more natural (and even mandatory) to use Bayesian updating, starting from a prior probability distribution that does not yet include the information about the average number of dots in $N$ throws. 
Importantly, this prior distribution would be over chances, and so would, in fact, correspond to $\Prt$.

The requirement of not including any information in the prior distribution, leads, when we stick to the same measure $m(\vec{c})$=constant as used in the preceding subsection, to 
$\Prt_{{\rm prior}}(\vec{c})=$C, with C a constant independent of $\vec{c}$ determined by normalization,
\be
{\rm C}=\frac{1}{\intdc \delta(\sum_n c_n-1)}.
\ee
One may, in fact, apply the MaxEnt procedure to obtain this particular prior using  normalization as the only constraint. (In a much more general context, one may, as a rule, use MaxEnt to set the prior distribution on the probabilities that underly one's problem: see, e.g., Caticha and Preuss (2004). In that case, one subsequently uses Bayesian updating to modify this distribution in the light of incoming empirical data.)

With this starting point
the Bayesian updating solution is constructed as follows.
First, as an intermediate step, assume for the moment that
the data given consisted of 
the numbers of times $M_n$ that $n$ spots were observed in $N$ tosses of the die, so that $\sum_n M_n=N$. In that case, the posterior distribution over chances would be determined by Bayes' rule as (cf.~Eq.~(\ref{DF}): the combinatorial pre-factor drops out here, but it will  reappear in (\ref{post}) below)
\be\label{MM}
\Prt'_{{\rm post}}(\vec{c})=
\frac{c_1^{M_1}\times c_2^{M_2}\ldots\times c_6^{M_6}}{\intdc c_1^{M_1}\times c_2^{M_2}\ldots\times c_6^{M_6}}.
\ee
Now the actual information provided in the Brandeis Dice Problem is (much) less than this, and we only know that
\be\label{cMM}
\sum_n nM_n/N=\expect{B}=4.5.
\ee
The posterior distribution is, in this case, a weighted
average over those distributions of the form (\ref{MM})
for which the constraint (\ref{cMM}) is satisfied. So, up to an overall normalization factor, we have 
\be\label{post}
\Prt_{{\rm post}}(\vec{c})\propto
\sum_{\substack{M_1,M_2,\ldots M_6\\
                 \sum_n nM_n/N=\expect{B}\\
                  \sum_n M_n=N}}
\frac{N!}{M_1!M_2!\ldots M_6!}c_1^{M_1}\times c_2^{M_2}\ldots\times c_6^{M_6}.
\ee
This solution can be evaluated numerically for small values of $N$ (Uffink 1996 presents results for several values of  $N$ up to 60). It differs from Jaynes' solution in general, even in the limit of $N\rightarrow\infty$. Just in case the reader is surprised by this, let me note that (i) Seidenfeld (1986) showed how updating and MaxEnt procedures lead to equivalent results only in very special circumstances (and the Brandeis Dice Problem does not constitute one of those special circumstances); (ii) Skyrms (1987)  argued that the MaxEnt procedure is not conceptually equal to updating and as such one should not expect their results to be the same.

The Bayesian updating solution actually becomes very simple in the limit of $N\rightarrow\infty$.\footnote{This limit is what I take to be the intention behind Jaynes' formulation of his toy problem (``...a very large number $N$ of times...'', see quote in the Introduction).} One gets (up to a normalization constant)
\be\label{Micro}
\Prt(\vec{c})\propto \delta(\sum_n nc_n-\expect{B})
\delta(\sum_n c_n-1).
\ee
The Bayesian updating solution for $P_n$ in the limit $N\rightarrow\infty$ is, therefore, 
\be\label{Bayes}
P_n=\frac{\intdc c_n\delta(\sum_n c_n-1)\delta(\sum_n nc_n-\expect{B})}{\intdc \delta(\sum_n c_n-1)\delta(\sum_n nc_n-\expect{B})}.
\ee
One may find  the numerical results for the case $\expect{B}=4.5$ in Table 2, where they are compared with Jaynes' original solution. 
\begin{table}[h]
\begin{center}
\begin{tabular}{|c|c|c|c|}
\hline
\hline
Probability&Jaynes& Bayes &(Error bar)\\
\hline
\hline
$P_1$&0.054&0.062& (0.050)\\
\hline
$P_2$&0.079&0.078& (0.062)\\
\hline
$P_3$&0.114&0.103 &(0.083)\\
\hline
$P_4$&0.165&0.153& (0.125) \\
\hline
$P_5$&0.240&0.262& (0.146)\\
\hline
$P_6$&0.347&0.342 &(0.196)\\
\hline
\hline
\end{tabular}
\end{center}
\label{default}
\caption{Jaynes' MaxEnt solution to the Brandeis Dice Problem (second column) compared with an alternative Bayesian updating solution (third column), described in Section \ref{BU}, for the limiting case where $N\rightarrow\infty$. The entry headed ``(Error bar)'' gives the standard deviation $\sigma_n$ in this alternative estimate of $P_n$, as calculated in Section \ref{SD}. }
\end{table}

For completeness, I note that in this particular case another interesting comparison is with a die
such that
the average number of dots shown was actually 3.5, as expected for a fair die.
As Uffink (1996) showed, 
the Bayesian updating solution for $\{P_n\}$ is {\em not} what one might naively have expected, namely, the uniform probability distribution (Jaynes' MaxEnt method does yield this uniform solution).
The solution is different, because the knowledge
we have is quite distinct in the following two cases:
(i) not knowing anything leads to the uniform distribution for $\Prt$, which leaves open the possibility that
the expected average number of dots is anywhere between 1 and 6, and (ii) 
knowing that the expected average number of dots in fact equals $\expect{B}=3.5$ eliminates almost all possible distributions $\Prt$.
In Table 3 I display the values of the probabilities in the limiting case $N\rightarrow \infty$ (Uffink provided values for several finite values of $N$ for this case, too).

\begin{table}[h]
\begin{center}
\begin{tabular}{|c|c|c|}
\hline
\hline
Probability&Bayes&(Error bar)\\
\hline
\hline
$P_1=P_6$&0.141& (0.095)\\
\hline
$P_2=P_5$&0.166& (0.120)\\
\hline
$P_3=P_4$&0.193&(0.153)\\
\hline
\hline
\end{tabular}
\end{center}
\label{default}
\caption{Bayesian updating solution for  $\expect{B}=3.5$ in the case $N\rightarrow\infty$, as calculated in Section \ref{BU}.}
\end{table}

\section{Error bars}\label{SD}
With the solutions based on $\Prt(\vec{c})$ in hand, we can do one more thing, namely, evaluate the standard deviations in the estimates for $P_n$, defined as 
\be
\sigma_n:=\sqrt{\intdc c_n^2 \Prt(\vec{c})-P_n^2}.
\ee
Tables 1 and 2 list these standard deviations for both alternative solutions constructed in Section 2. One sees that $\sigma_n$ is quite large, of the same order of magnitude as $P_n$ itself. 

The large error bars are the reason for displaying fewer digits in Tables 1 and 2 than Jaynes originally did (he produced 5 digits), so as to avoid the misleading impression of high precision.
For the same reason, it seems somewhat futile to argue about, say, whether Jaynes' assignment of $34.7\%$ to $P_6$ is more reasonable than the alternative assignments of $P_6=38\%$ or $P_6=34.2\%$, respectively, given that the error bar in the latter two (Jaynes' estimate came without an error bar!) is about $20\%$ so that these 
 results are not statistically significantly different.

Let me now discuss why error bars for probability estimates are useful. First of all, I suppose it goes almost without saying that, in the case of physical quantities, specifying just the average observed value is in general not sufficient to yield reliable knowledge. A well-known case in point is provided by the story of the man who drowned in a river that he knew to be 1m deep on average.

The quantities we are interested in here, however, are (subjective) probabilities, not physical quantities.
Such probabilities are relied upon when betting.
If one places just a single bet on the occurrence of one particular event, then all one needs to know is one's degree of belief in that occurrence. The standard deviation does not matter, and acceptable odds on the bet are determined entirely by one's degree of belief.
However, 
as soon as one places a bet on {\em two} events of the same type, the standard deviation does play a role. In fact, if one bets on the occurrence of 
two subsequent identical outcomes $n$ in our dice experiment, then the correct probability  assigned  to that combination of events is
\be
P_{nn}=\intdc \Prt(\vec{c})c_n^2=P_n^2+\sigma_n^2,
\ee
and not simply $P_n^2$. It is clear, furthermore, that one would need to specify the covariance matrix $P_{nm}=\intdc \Prt(\vec{c})c_nc_m$ for all combinations $(n,m)$ if one were to place bets on all possible pairs of outcomes. (And, of course, knowing $\Prt$ allows one to calculate higher-order moments, namely, those necessary to determine acceptable odds for bets on more than two events.)

Interestingly, Jaynes discusses in  his 1978 overview paper the probability one ought to assign to the occurrence of two particular events in a row. He admits he needs the covariance matrix $P_{nm}$, but he cannot obtain it from his solution.
Jaynes states that his ``maximum-entropy solution does not, and cannot, make any statement about frequencies'' and that it applies only to predictions of the {\em single} toss of the die.
He subsequently does proceed to provide a possible range of values for $P_{nm}$ based on the assumption that the tosses of the die form an exchangeable sequence, but he stops short of applying the De Finetti theorem.

\section{Comparison with Statistical Mechanics}\label{Stat Mech}
Eq.~(\ref{Micro}) shows that Bayesian updating leads to a mathematical analogue of the micro-canonical ensemble, whereas Eq.~(\ref{Can1}) 
shows that the MaxEnt procedure applied at the same level of description leads to a mathematical analogue of the canonical ensemble. In this Section I will consider these purely mathematical analogies in more detail. Subsequently I will discuss some essential disanalogies, concerning physical aspects, between the Brandeis Dice Problem and Statistical Mechanics.

\subsection{Canonical ensemble analogies}
Here I compare the two different MaxEnt solutions of the Brandeis Dice Problem with two different physical canonical ensembles, describing two different physical systems.
In Table 4 I list the microscopic variables, the measure on the space of those variables, and the Boltzmann factor for these two physical systems.
\begin{table}[h]
\begin{center}
\begin{tabular}{|c|c|c|c|c|}
\hline
\hline
&Stat Mech$^a$&Stat Mech$^b$&Chances&Jaynes\\
\hline
\hline
micro variables&$\vec{r}_n,\vec{p}_n$&$|n\rangle$&$\vec{c}_n$&$n$\\
\hline
range of $n$&$n=1\ldots N_p$&$n=1\ldots\infty$&$n=1\ldots 6$&$n=1\ldots 6$
\\
\hline
measure&$\Pi_n\intdrp/h^{3N}$&$\sum_n$&$\Pi_n\intdc\!_n$&$\sum_n$\\
\hline
Boltzmann factor&$\exp[-\beta \sum_n \vec{p}_n^2/(2m_n)]$&$\exp(-\beta \alpha n^2)$&$\exp[-\lambda
\sum_n nc_n]$&$\exp[-\lambda n]$\\
\hline
\end{tabular}
\end{center}
\label{default}
\caption{Four canonical ensembles: Stat Mech$^a$ refers to a gas of $N_p$ classical distinguishable noninteracting particles, Stat Mech$^b$ refers to a single quantum particle of mass $m$ in a 1D infinite square well of length $L$. The constant $\alpha$ has the value $\alpha=h^2/(8mL^2)$ with $h$ Planck's constant. The constant $\beta$, as usual, stands for $\beta=1/(k_BT)$, with $k_B$ Boltzmann's constant and $T$ the temperature. $\beta$ may be interpreted as a Lagrange multiplier, just like $\lambda$ in the solutions to the dice problem. One notices the close analogies between the entries in the ``Stat Mech$^a$'' and ``Chances'' columns, as well as 
those between the  ``Stat Mech$^b$'' and ``Jaynes'' columns.
}
\end{table}
The first system is an ideal gas of $N_p$ classical distinguishable particles. The microscopic variables
are continuous and are chosen to be canonically conjugated pairs. I picked position and momentum, but there are alternatives, in particular, action-angle variables. The measure on the space is not actually determined by classical physics, but it follows from quantum mechanics that the correct measure is $\intdrph$ for each particle, with $h$ Planck's constant.
This case is manifestly analogous to 
the alternative MaxEnt solution: the microscopic variables are continuous, and they are enumerated by a discrete variable $n$, which runs to $N_p$ in the physical case, and to $M=6$ in the dice case. The measure
is an $N_p$-fold integral or an $M$-fold integral over the continuous variables. Thus, the number of particles, $N_p$, is analogous to the number of sides of the die, $M$.

The second physical situation concerns a single spinless quantum particle in a 1D infinite square well. The one (discrete) variable describing the state is the quantum number $n$ for energy.
Jaynes' MaxEnt solution is mathematically analogous to this case. The microscopic variable is discrete, and the measure is simply a sum over all $n$. One difference is that in the quantum case the range of values of the integer $n$ is unlimited (even for a single particle), while in the die case it is limited to $n=1...M$, but this difference could be eliminated by simply (even if artificially) imposing an upper limit to the energy of the quantum particle.

\subsection{The thermodynamic limit}
The above identifications allow us to
find the analogue of the thermodynamic limit.
Since for the classical ideal gas  this limit corresponds to the limit $N_p\rightarrow\infty$,
the straightforward analogue  for the alternative
MaxEnt solution is the limit $M\rightarrow\infty$ for $M$-sided dice. 

To see this in more detail, it is illuminating to consider
the equivalence, within Stat Mech, of the canonical and micro-canonical ensembles in the thermodynamic limit. The reason for this equivalence is that
 the density of states, $\rho(E)$, tends to increase monotonically with energy $E$, so that the probability density for finding a state with a particular energy in the canonical ensemble (which is the product of the density $\rho(E)$ with the decaying Boltzmann factor $\exp(-\beta E)$)
has a maximum around the mean energy $\expect{E}$. This maximum becomes more pronounced with increasing number of degrees of freedom and in the thermodynamic limit the canonical distribution becomes a delta function, hence equal to the micro-canonical distribution $\delta(E-\expect{E})$.

For the Brandeis Dice Problem, 
let us define the analogue of $E$ as
\be
B:=\sum_{n=1}^M nc_n,
\ee
in terms of the chances $c_n$ and the number of sides, $M=6$.
The analogue of the density of states, denoted by $\rho(B)$, is then the number of solutions for $\vec{c}$ given a value of $B$.
This function is {\em not} a monotonically increasing  function of $B$. Instead, it reaches a maximum for the ``honest value'', $B=(1+6)/2=3.5$. In particular, it is a decreasing function of $B$ for $B>3.5$, and an increasing function for $B<3.5$.
As noted before, the Lagrange multiplier $\lambda$ is negative for $\expect{B}>3.5$ and positive for
$\expect{B}<3.5$.
In either case, it follows that the ``canonical distribution'' $\rho(B)\exp(-\lambda B)$ will reach a maximum for a value of $B$ around $\expect{B}$. But this maximum is by no means infinitely sharp for a standard die.  The ``canonical ensemble'' (obtained by the MaxEnt procedure) does not, therefore, reduce to the ``micro-canonical ensemble'' obtained by Bayesian updating. The reason is that the number of degrees of freedom is so small: there are just 5 independent chances $c_n$.  In the limit $M\rightarrow\infty$, however, the maximum becomes sharp.

This analogy between the number of gas particle and the number of sides of the die also explains why the error bars
$\sigma_n$ in the estimates for $P_n$ are so large, relatively speaking.
For a macroscopic gas,
the fluctuations of the physical state variables around thermal equilibrium
are of relative size $1/\sqrt{N_p}$.
These relative fluctuations, therefore, go to zero in the thermodynamic limit and are very small for typical macroscopic numbers (say, $N_p\approx 10^{23}$). For the dice problem, in contrast, $M$ is by no means large and $1/\sqrt{M}$ is appreciable.

\subsection{Disanalogies}
The analogies discussed above are all of a mathematical sort.
Physically, there are certain obvious differences between the dice problem and typical Statistical Mechanics problems. These differences may teach us something about the relation between Stat Mech and MaxEnt.
Here is a small list:

First, energy conservation plays a crucial role in Stat Mech. We arrive at the canonical and micro-canonical ensembles by imposing conditions on the average or the total energy of a system, respectively. It makes sense to do that {\em only} because energy is conserved. 
Imagine, in contrast, that we somehow managed to measure the quantity $A:=\sum_n \vec{p}_n^4$ of a gas at some instant of time. After one collision, this quantity would already have a different value. On an extremely short time scale, therefore, our knowledge of the value of $A$ would become obsolete---unless we could follow the precise microscopic dynamics and thus keep track of the rapidly changing value of $A$--- and we could not apply the corresponding micro-canonical ensemble. The Brandeis Dice Problem has no conservation law (of the number of spots, for instance). Rather, by construction, the toy problem has no dynamics at all as soon as we assume exchangeability to hold for the sequence of tosses. 

Second, for Statistical Mechanical problems the appropriate settings for the two ensembles are clear: one uses the micro-canonical ensemble for an isolated system (whose energy is conserved) and the
canonical ensemble is used for a small system in thermal contact with a large reservoir, such that reservoir plus small system are isolated. There are no analogues of reservoirs for the Brandeis Dice Problem, nor of the exchange of energy. Rather, as we have seen, Bayesian updating leads to the micro-canonical ensemble whereas MaxEnt leads to the canonical ensemble.

Third, for the dice problem, there is a choice on what level of description the MaxEnt procedure is to be applied, and that is why one may construct (at least) two canonical ensembles. Moreover, even if we have settled on the probability distribution $\Prt(\vec{c})$ as the appropriate level of description, there is still the choice of the measure $m(\vec{c})$.
The Statistical Mechanical canonical ensemble of a given physical system, on the other hand, is unique, 
because the appropriate way of counting states is uniquely determined by the laws of quantum mechanics: one uses the dimension of the Hilbert space spanned by states of a given energy. This sets the appropriate level of description as well as which measure to use.

Finally, there is yet another way to 
treat dice problems, namely, by treating the dice as physical systems.
We could, in principle, consider all relevant microscopic physical variables that
determine the motion of the die as it is tossed. This may include the positions and momenta of  the ambient air molecules, the condition of the surface on which the die will fall, and so on and so forth. 
By doing this we would have 
more than just a mathematical analogy, but we also would have
lost sight of the fact that we were merely trying to pose and solve an illustrative toy problem. Conversely,\footnote{As pointed out by one of the referees.} we could enhance the analogy  
by treating Stat Mech problems in a nonstandard way: we could introduce chances
for the physical system to be in given microstates, and then introduce degrees of belief about those chances.

\section{Conclusions}
I presented two solutions to the Brandeis Dice Problem that differ from Jaynes' original solution. One used the MaxEnt procedure (but at a different level of description than Jaynes did), the other Bayesian updating. 
The former gave rise to a mathematical analogue of the canonical ensemble, the latter to an analogue of the micro-canonical ensemble.
  
The Bayesian updating and MaxEnt solutions (and thereby the analogues of the micro-canonical and canonical ensembles) are {\em not} equivalent, but I showed how they become equivalent in what would have to be the analogue of the thermodynamic limit, namely the Brandeis Dice Problem generalized to $M$-sided dice with $M\rightarrow\infty$.

The main advantage of the two new solutions is that they automatically come equipped with error bars. These error bars for the dice problem are large and they remind us that lack of knowledge, as encoded by the MaxEnt principle, does not lead to {\em precise} estimates of probabilities. Moreover, such error bars are necessary for the calculation of joint probabilities for the outcomes of two  dice tosses. Jaynes had explicitly noted that his solution does not and cannot answer questions about joint probabilities.

Purely physical considerations, such as conservation of energy and the existence of thermal reservoirs, play no role in the Brandeis Dice Problem, but are crucial in standard Statistical Mechanics. This shows in what sense Stat Mech is more than just an application of MaxEnt.

\section*{Acknowledgements}
I thank both anonymous referees for their extensive and incisive comments on an earlier version of this paper.

\section*{References}

\begin{itemize}
\item Caticha, Ariel, and Preuss, Roland. ``Maximum entropy and Bayesian data analysis: Entropic prior distributions." Physical Review E 70.4 (2004): 046127.

\item De Finetti, Bruno. ``Theory of Probability: A critical introductory treatment.'' Vol. 1. Wiley, 1974.

\item Friedman, Kenneth, and Abner Shimony. ``Jaynes's maximum entropy prescription and probability theory." Journal of Statistical Physics 3.4 (1971): 381-384.

\item Jaynes, Edwin T. ``Information theory and statistical mechanics." Physical Review 106.4 (1957): 620.

\item Jaynes, Edwin T. ``Information Theory and Statistical Mechanics (Notes by the lecturer).'' 
Statistical Physics 3, Lectures from Brandeis Summer Institute 1962. New York: WA Benjamin, Inc., 1963., p.181.

\item Jaynes, Edwin T. ``Where do we stand on maximum entropy." The maximum entropy formalism (1978): 15-118.

\item Jaynes, Edwin T. ``Some random observations." Synthese 63.1 (1985): 115-138.

\item Kyburg Jr, Henry E. ``Higher order probabilities and intervals." Int. J. Approx. Reasoning 2.3 (1988): 195-209.

\item Maher, Patrick. ``What is probability." Unpublished manuscript.\\
 {\tt http://patrick.maher1.net/preprints/pop.pdf} (2010).

\item Porto Mana, P. G. L. ``On the relation between plausibility logic and the maximum-entropy principle: a numerical study." arXiv preprint arXiv:0911.2197 (2009). 
 
\item  Rowlinson, J. S. ``Probability, information and entropy." Nature 225 (1970): 1196-1198.

\item Seidenfeld, Teddy. ``Entropy and uncertainty." Advances in the Statistical Sciences: Foundations of Statistical Inference. Springer Netherlands, 1986. 259-287.

\item Skyrms, Brian. ``Maximum entropy inference as a special case of conditionalization." Synthese 63.1 (1985): 55-74.

\item Skyrms, Brian. ``Updating, supposing, and MaxEnt." Theory and Decision 22.3 (1987): 225-246.

\item Uffink, Jos. ``The constraint rule of the maximum entropy principle." Studies in History and Philosophy of Science Part B: Studies in History and Philosophy of Modern Physics 27.1 (1996): 47-79.

\item Zabell, Sandy L. ``The rule of succession." Erkenntnis 31.2-3 (1989): 283-321.

\end{itemize}
\newpage
\section*{Appendix: Laplace's Rule of Succession}
This Appendix provides error bars  for Laplace's ``Rule of Succession.'' This rule concerns a situation in which there are $M$ possible outcomes of an experiment, and one has observed one particular outcome $k$ times out of $m$ cases in total. Then, according to the Rule of Succession, the probability assigned to the next case having that outcome should be $P_{k;m}=(k+1)/(m+M)$. This  includes as a special case the assignment of probability $P_{0,0}=1/6$ to $n$ spots (for any $n=1\ldots 6$)  coming up on the throw of a die for which we know nothing except that the die has $M=6$ sides, one if which is guaranteed to come up.
Interestingly, the Rule of Succession is equivalent to
assuming a flat distribution for $\Prt$.
That is, not only does the Rule of Succession follow (straightforwardly) from a flat distribution, the converse statement, that a flat distribution for $\Prt$ is implied by the Rule of Succession, is true, too (see Zabell 1989).

Thanks to this equivalence, the standard deviation in the estimate 1/6 for an unknown six-sided die to come up with any particular side follows immediately from the Rule of Succession itself (since the probability to see $n$ spots turn up on the second toss given that the first toss produced $n$ spots is 2/7):
\be
\sigma^2=\frac{1}{6}\times\frac{2}{7}-\left(\frac{1}{6}\right)^2=\frac{5}{7}\times\frac{1}{36},
\ee
so that
\be
\sigma=\frac{\sqrt{35}}{42}\approx 0.141.
\ee
Again one sees that the standard deviation is almost as large as the probability, 1/6, itself.
For completeness, I note that the standard deviation in the
general probability estimate of $P_{k;m}$ with the same method is found to be
\be
\sigma^2_{k;m}=\frac{(k+1)(M+m-k-1)}{(m+M)^2(m+M+1)}.
\ee

\end{document}